\documentclass[traditabstract]{aa} 
\usepackage{graphicx}
\usepackage{natbib}
\usepackage{mwe,tikz}
\usepackage[percent]{overpic}
\usepackage{txfonts}
\usepackage{hyperref}
\def\igr{Swift\,J1734.5-3027}
\def\inte{{\em INTEGRAL}}
\def\xmm{{\em XMM-Newton}}

\def\swift{{\em Swift}}

\def \inte {{\em INTEGRAL}}
\def \xmm {{\em XMM--Newton}}

\def \ergsec{\hbox{erg s$^{-1}$}}
\def \ferg {erg cm$^{-2}$ s$^{-1}$}
\def \hcm {\hbox {\ifmmode $ atom cm$^{-2}\else atom cm$^{-2}$\fi}}

\def \arcsec {\hbox{$^{\prime\prime}$}}

\def \apjl {ApJL}
\def \apjs {ApJS}\def \aap {A\&A}

\begin{document}
   \title{Swift\,J1734.5-3027: a new long type-I X-ray bursting source}

   \author{E. Bozzo
    \inst{1}
   \and P. Romano
     \inst{2}
   \and M. Falanga 
      \inst{3,4}   
   \and C. Ferrigno 
     \inst{1}
   \and A. Papitto
   \inst{5}  
   \and  
   H.A.\ Krimm
   \inst{6,7}  
    }

   \institute{ISDC Data Centre for Astrophysics, Chemin d'Ecogia 16,
    CH-1290 Versoix, Switzerland; \email{enrico.bozzo@unige.ch}
    \and 
    INAF, Istituto di Astrofisica Spaziale e Fisica Cosmica - Palermo, via U. La Malfa 153, 90146 Palermo, Italy
     \and 
     International Space Science Institute, Hallerstrasse 6, CH-3012 Bern, Switzerland
    \and 
     International Space Science Institute in Beijing, No.~1 Nan Er Tiao, Zhong Guan Cun, Beijing 100190, China
     \and 
     Institut de Ci\`encies de l'Espai (IEEC-CSIC), Campus UAB, carrer de Can Magrans, S/N 08193, Cerdanyola del Vall\`es, Barcelona, Spain 
     \and 
      Center for Research and Exploration in Space Science and Technology (CRESST) and NASA Goddard Space Flight Center, Greenbelt, MD 20771  USA 
      \and
      Universities Space Research Association, 7178 Columbia Gateway Drive, Columbia, MD 21046  USA 
     }

   \date{}

  \abstract{Swift\,J1734.5-3027 is a hard X-ray transient discovered by \swift\ while undergoing an outburst in September 2013. 
  Archival observations showed that this source underwent a previous episode of enhanced X-ray activity in May-June 2013. 
  In this paper we report on the analysis of all X-ray data collected during the outburst in September 2013, the first that could 
  be intensively followed-up  by several X-ray facilities. 
  Our data-set includes \inte,\ \swift,\ and \xmm\ observations. From the timing and spectral analysis of these 
  observations, we show that a long type-I X-ray burst took place during the source outburst, making Swift\,J1734.5-3027 a new member of the 
  class of bursting neutron star low-mass X-ray binaries. The burst lasted for about 1.9~ks and reached a peak flux of 
  (6.0$\pm$1.8)$\times$10$^{-8}$~erg~cm$^{-2}$~s$^{-1}$ in the 0.5-100~keV energy range. The estimated burst fluence in the same energy range 
  is (1.10$\pm$0.10)$\times$10$^{-5}$~erg~cm$^{-2}$. By assuming that a photospheric radius expansion took place 
  during the first $\sim$200~s of the burst and that the accreted material was predominantly composed by He, we derived a distance 
  to the source of 7.2$\pm$1.5~kpc.}   
  

   \maketitle

\section{Introduction}
\label{sec:intro}

Swift\,J1734.5-3027 is a hard X-ray transient discovered by \swift\ on 2013 September 1 \citep{malesani13}. 
At discovery, the source spectrum could be reasonably well described in the \swift\,/BAT energy band 
(15-150~keV) by either a blackbody model with temperature $kT$=2.5$\pm$0.4~keV, or a power-law model with 
photon index $\Gamma$=5.8$\pm$0.7. A fluence of (1.6$\pm$0.3)$\times$10$^{-7}$~erg~cm$^{-2}$ was recorded during the 
initial $\sim$27~s of the BAT trigger \citep{kennea13}. These spectral characteristics led \citet{kennea13} to suggest 
that \igr\ was a new superbursting X-ray transient. 
The source outburst was also observed with the instruments on-board \inte,\  
providing additional information on the source broad-band spectrum \citep{kuulkers13}. 
An archival search into previous \swift\ observations performed in the directions of \igr\ showed that the source  
underwent a likely previous outburst in May-June 2013 \citep{laparola13}. 
Unfortunately, in that occasion no follow-up observations were performed. 

In this paper, we report on all X-ray data collected from \igr\ during its outburst in September 2013. Beside the data from the 
\inte\ and \swift\ monitoring, we also analyze a Target of Opportunity observation that we obtained with \xmm\ on 2013 September 8. 
From the timing and spectral analysis of these data, we show that Swift\,J1734.5-3027 is a new 
member of the transient bursting neutron star low-mass X-ray binaries (NSLMXBs) and that the BAT discovery of the source occurred 
during a long type-I X-ray burst (rather than a superburst as suggested previously).  
\begin{figure}
  \includegraphics[width=8.5cm]{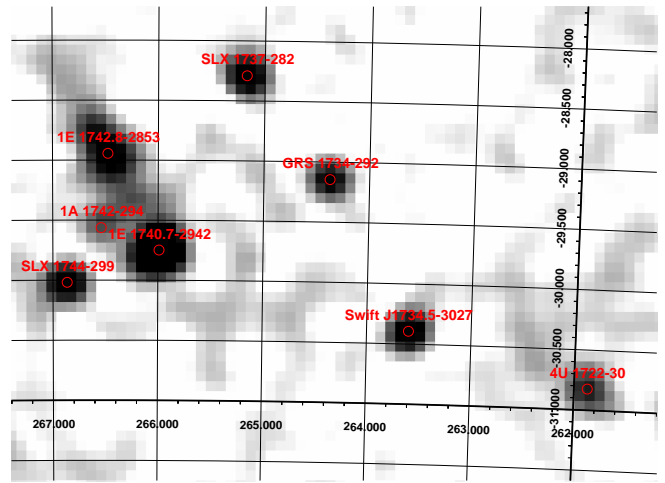}
  \caption{\inte\ IBIS/ISGRI mosaiked image centered around the position of Swift\,J1734.5-3027. We used all available ISGRI data 
  mentioned in Table~\ref{tab:integral} to build the mosaic (20-70~keV). The source is detected in this mosaic 
  with a significance of 34$\sigma$.}  
  \label{fig:mosa}
\end{figure}

\section{\inte\ data}
\label{sec:integral}

\igr\ was detected by \inte\ for the first time during the observations performed toward the   
Galactic bulge in the satellite revolution 1329, i.e. about half a day before the BAT discovery 
(from 56535.85950~MJD to 56536.01338~MJD; see Table~\ref{tab:integral}). It remained within the FoV of the instruments on-board \inte\ 
until satellite revolution 1348 (from 56592.54664~MJD to 56593.31612~MJD), when the window of seasonal visibility toward the Galactic Center 
closed. We analyzed all \inte\ data by using version 10.1 of the Off-line Scientific Analysis software   
(OSA) distributed by the ISDC \citep{courvoisier03}. \inte\ observations are divided into ``science windows'' (SCWs), 
i.e. pointings with typical durations of $\sim$2-3~ks. 
To limit the ISGRI calibration uncertainties \citep{lebrun03}, we made use of all public SCWs 
from the Galactic Bulge, Scutum/Sagittarium, and Perseus/Norma monitoring programs during which 
the source was located to within 12~deg from the center of the IBIS FoV \citep{ubertini03}.  
We also included in our dataset the observations of the Galactic Center and 4U\,1728-34 
for which our group was awarded data rights in revolutions 1329-1348. 
A summary of the total exposure-time available in each revolution is provided in Table~\ref{tab:integral} for IBIS/ISGRI  
and the two JEM-X telescopes \citep{lund03}. We extracted the mosaics and spectra in each revolution for the two JEM-X and ISGRI. 
All JEM-X spectra were extracted by using the standard 16-channels response matrix, while a customized 37 energy bin response 
matrix was created for ISGRI in order to optimize the signal-to-noise ratio (S/N) in the energy range (20-50\,keV). 
Only in revolution 1329, the ISGRI spectrum 
was extracted with a reduced energy binning (8 channels), as the source was relatively faint (see Sect.~\ref{sec:swift}).  
JEM-X lightcurves with a time resolution of 2~s were extracted from all observations to search for type-I X-ray bursts, 
but none was found. 
We did not perform other timing analyses of the \inte\ data as the source was too faint to extract meaningful power spectra. 
In Fig.~\ref{fig:mosa} we show the ISGRI mosaic realized by using all available data. 
\begin{table*}
\centering
\scriptsize
\caption{Log of the \inte\ data used in this paper. We report the effective exposure times, detection significance and estimated 
fluxes for ISGRI, JEM-X1, and JEM-X2 in all revolutions considered for the data analysis. The detection significances are reported 
for ISGRI (JEM-X1,2) in the 20-40~keV (3-10~keV) energy band.} 
\begin{tabular}{@{}lllllll@{}}
\hline
\hline
\noalign{\smallskip}
Rev.  & \multicolumn{2}{c}{Observing time} & \multicolumn{2}{c}{Exposure time} & \multicolumn{2}{c}{Flux$^b$} \\
   & START & STOP & ISGRI & JEM-X$^b$  & ISGRI & JEM-X$^c$ \\
   & (MJD)   &   (MJD)   & (ks) & (ks) & (20-70~keV) & (3-35~keV) \\
\noalign{\smallskip}
\hline
\noalign{\smallskip}
1329  & 56535.85950 & 56536.01338 & 8.6  & 4.9 & 13.0$\pm$2.0 & $<$12.0 \\
\noalign{\smallskip}
1330  & 56538.69428 & 56539.63178 & 73.3  & 8.0 & 19.7$\pm$1.0 & 16.1$\pm$4.0\\
\noalign{\smallskip}
1331  & 56541.84263 & 56542.66694 & 63.2  & 3.2 & $<$20.5 & $<$3.5 \\
\noalign{\smallskip}
1332  & 56544.85712 & 56545.84130 & 74.0  & 7.7 & $<$2.7 & $<$9.8\\
\noalign{\smallskip}
1333  & 56548.63922 & 56548.79324 & 10.9  & 4.7 & $<$3.8 & $<$12.0\\
\noalign{\smallskip}
1334  & 56552.33453 & 56553.07995 & 59.3  & 7.3 & $<$3.1 & $<$12.5\\
\noalign{\smallskip}
1335  & 56553.63791 & 56554.41490 & 58.0  & 10.8 & $<$3.0 & $<$8.7 \\
\noalign{\smallskip}
1336  & 56556.62811 & 56559.26417 & 71.3  & 14.6 & $<$2.3 & $<$7.2 \\
\noalign{\smallskip}
1337  & 56559.80640 & 56560.39413 & 45.4  & 10.9 & $<$2.9 & $<$8.2 \\
\noalign{\smallskip}
1339  & 56565.59969 & 56568.21845 & 199.7  & 23.9 & 8.8$\pm$0.4 & 7.2$\pm$2.0 \\
\noalign{\smallskip}
1340  & 56569.43604 & 56571.05972 & 105.2  & 25.8 & 7.6$\pm$0.7 & $<$5.8 \\
\noalign{\smallskip}
1341  & 56572.63094 & 56574.06932 & 112.8  & 24.5  & 5.1$\pm$0.7 & $<$5.6 \\
\noalign{\smallskip}
1343  & 56577.59515 & 56578.53990 & 69.0  & 13.6 & 9.8$\pm$0.9 & $<$7.6 \\
\noalign{\smallskip}
1344  & 56580.71753 & 56582.17201 & 109.7  & 25.7 & 7.5$\pm$0.6 & $<$5.5 \\
\noalign{\smallskip}
1345  & 56586.02326 & 56586.58723 & 45.5  & 7.6 & 14.0$\pm$1.0 & 15.3$\pm$3.6 \\
\noalign{\smallskip}
1346  & 56586.73297 & 56589.00642 & 110.4 & 20.1 & 8.7$\pm$0.7 & 10.8$\pm$2.2 \\
\noalign{\smallskip}
1347  & 56589.72509 & 56590.51950 & 63.3  & 12.0 & 6.6$\pm$0.8 & 19.0$\pm$2.8 \\
\noalign{\smallskip}
1348  & 56592.54664 & 56593.31612 & 60.8  & 5.8 & 8.3$\pm$1.1 & 22.2$\pm$5.0 \\
\noalign{\smallskip}
\hline
\noalign{\smallskip}
\end{tabular}
\tablefoot{$^a$: We considered a detection reliable in ISGRI and JEM-X only $>$6$\sigma$ and $>$4$\sigma$, respectively. 
$^b$: The flux in mCrab has been estimated by comparing the source count-rate measured from the mosaic 
to the count-rate from the Crab in the same energy band. The count-rate from the Crab has been measured 
by extracting ISGRI, JEM-X1, and JEM-X2 mosaics of the public calibration observations performed in satellite revolutions 
1342\footnotetext{See http://integral.esac.esa.int/isocweb/schedule.html?selectMode=rev\&action=schedule\&startRevno=1342\&endRevno=1342} 
(i.e. the calibration observations performed during the outburst of \igr). From these data we measured a Crab count-rate of 
206.1$\pm$0.2~cts~s$^{-1}$ in ISGRI (20-70 keV energy band) and 119.4$\pm$0.6~cts~s$^{-1}$ (134.5$\pm$0.7) for 
JEM-X2 (JEM-X1) in the 3-35~keV energy band.  
$^c$: We used JEM-X2 as a reference. 
Lower significance detections have been considered only to derive upper limits on the source X-ray emission 
(all upper limits are at 3$\sigma$ c.l.).}
\label{tab:integral}
\end{table*} 
\begin{figure}
  \includegraphics[width=6.2cm,angle=-90]{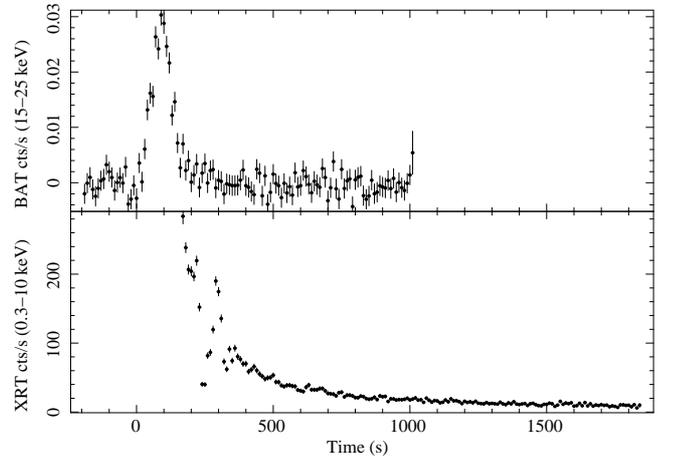}
  \caption{The BAT lightcurve corresponding to the on-board trigger at the onset of the outburst 
  from \igr\ (upper panel), together with the lightcurve extracted from the first available XRT observation 
  performed after the \swift\ automatic slew in the direction of the source (ID.~00569022000). The time bin of both curves is 
  10~s. The start time ($t=0$) is set at 56536.38458~MJD.}   
  \label{fig:trigger}
\end{figure}
\begin{figure}
  \includegraphics[width=5.5cm,angle=-90]{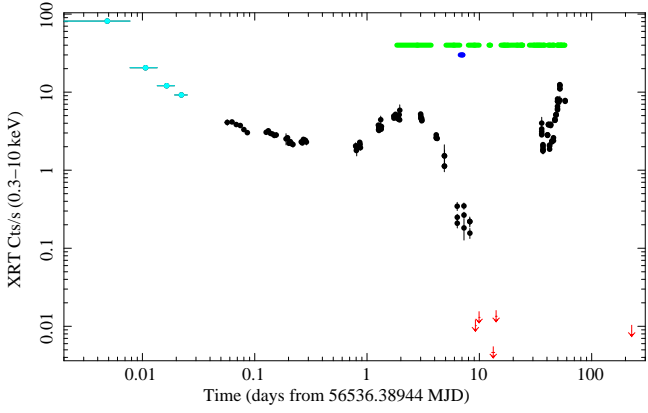}
  \caption{XRT lightcurve of the entire outburst observed by \igr\ (the start time is set 
  at 56536.38458~MJD as in Fig.~\ref{fig:trigger} and the bin time is 500~s). The points in cyan represents the long type-I 
  X-ray burst. The red arrows show the XRT observations in which the source was 
  not detected, providing an upper limit on the source count-rate. The green bands indicate the coverage of the outburst provided by 
  \inte\ (data of revolution 1329 are not shown, as they were collected before the \swift\ discovery). 
  We also indicated with a blue bar the time of the \xmm\ ToO observation.}  
  \label{fig:fig_xrtlc}
\end{figure}

\section{\swift\ data}
\label{sec:swift}

\subsection{XRT and BAT}
\label{sec:XRT}

\swift\,/XRT observations were carried out from 56536.38578~MJD to 56763.42292~MJD 
(see Table~\ref{tab:swift}). 
XRT data were collected in both windowed-timing (WT) and photon-counting (PC) 
mode and analyzed by using standard procedures within FTOOLS (v6.16).   
The XRT data were processed with the {\sc xrtpipeline} 
(v.0.13.1) and the latest calibration files available (20140730). 
PC source events were extracted in annular regions with 
outer radii ranging between 20 to 30 pixels  (1 pixel $\sim2.36$\arcsec)  
depending on source intensity. When required, the PC data were corrected for pile-up.  
We determined the size of the point spread function (PSF) core affected by pile-up 
through a comparison between the observed and nominal PSF \citep[see][]{vaughan2006:050315}.  
This sets the inner radius of the extraction region.  
WT source events were extracted in a circular region (20 pix radius), with the exception of 
observation 00569022000 during which the data were also affected by pileup; in the latter case 
an annular region (radii 2 and 20 pix) was adopted \citep[see][]{Romano2006:060124}. 
Background events were extracted in a circular regions with a radius of 70 pix for PC data 
and 20 pix for WT data. 
The {\sc xrtlccorr} task was then used to account pile-up, PSF losses, and
vignetting, in the light curves.  All light curves were background subtracted.  
For our timing analysis we converted the event arrival times to the 
solar system barycentre with {\sc barycorr}. 
For our spectral analysis, we extracted events in the same regions as those adopted 
for the light-curve creation; ancillary response files were generated with {\sc xrtmkarf}, 
to account for different extraction regions, vignetting and PSF corrections. 
The data were rebinned with a minimum of 20 count per energy bin to allow $\chi^2$ fitting.  
We also made use of BAT data corresponding to the on-board trigger of the source. 
The BAT lightcurve and spectra were extracted by using the {\sc batgrbproduct} tool available within 
{\sc Heasoft v. 6.16}. 

The best determined source position with XRT is found at: 
RA(J$2000)$ = $17^{\rm h} 34^{\rm m} 24\fs18$,
Dec(J$2000)=-30^{\circ} 23^{\prime} 53\farcs0$
with an associated uncertainty of 1\farcs4 (90\,\% c.l.).
This was obtained by using the online \swift\ ``User Object'' 
tools\footnote{\href{http://www.swift.ac.uk/user_objects/}{http://www.swift.ac.uk/user\_objects/}} , which 
uses field stars in the UVOT to determine the spacecraft attitude \citep{evans09}. 
\begin{figure}
  \includegraphics[width=6.25cm,angle=-90]{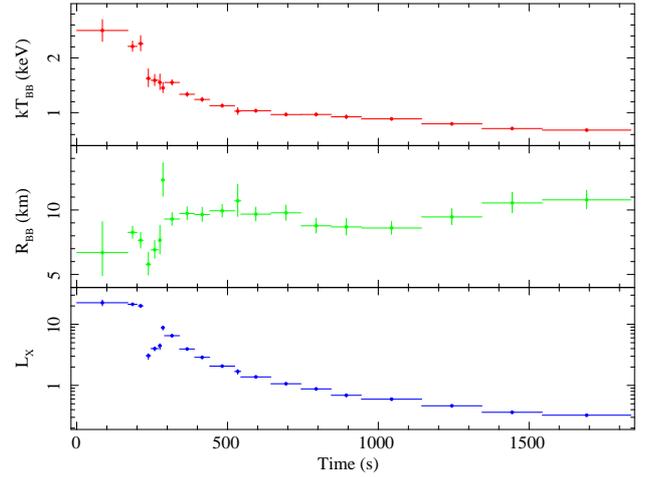}
  \caption{The results of the time resolved spectral analysis carried out during the first part of the 
  outburst decay. All spectra were fit with an absorbed blackbody model. We show in the three panels, from top to bottom,  
  the blackbody temperature, radius, and the X-ray luminosity in units of 10$^{37}$~erg~s$^{-1}$ 
  for an assumed distance of 7.2~kpc (see Sect.~\ref{sec:burst}).    
  The luminosity has been computed from the source flux extrapolated in the 0.5-100~keV energy band 
  (i.e., virtually bolometric for a blackbody spectral model with a temperature of $\sim$3~keV) and corrected for absorption. 
  The first point on the left is obtained from the fit to the BAT spectrum (in this case we also extrapolated the source flux 
  in the 0.5-100 keV energy band; see Sect.~\ref{sec:burst}).}    
  \label{fig:fig_xrtlc_zoom}
\end{figure}

The BAT lightcurve corresponding to the on-board trigger is shown in Fig.~\ref{fig:trigger}, together with the 
first XRT follow-up observation performed after the automatic slew of the spacecraft (first orbit of the observation ID.~00569022000). 
The XRT lightcurve covering the entire source outburst is shown in Fig.~\ref{fig:fig_xrtlc}. 
We notice that the source rose in flux rapidly at the onset of the event and then decayed slowly  
up to 56545~MJD. From 56545~MJD to 56571~MJD the source remained in a relatively low flux level  
before raising up again in X-rays on 56572~MJD. 
During the low luminosity interval, an \xmm\ ToO observation was carried out (see Sect.~\ref{sec:xmm}). 

\begin{figure}
  \includegraphics[width=6.3cm,angle=-90]{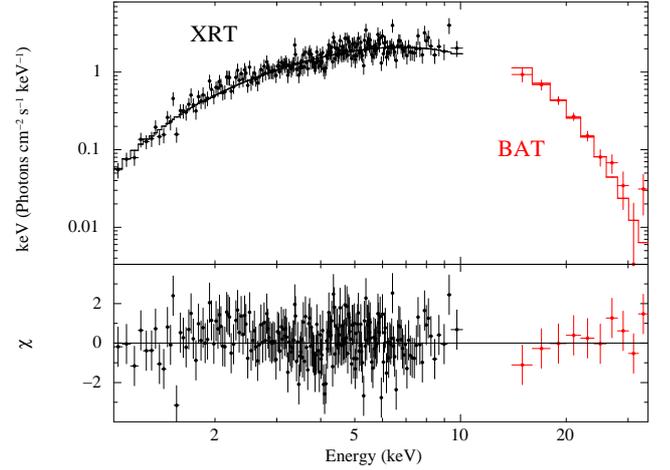}
  \caption{The quasi-simultaneous XRT and BAT spectrum of the source at the onset of the outburst. The BAT spectrum is integrated 
  for 113~s around the peak of the event, while the XRT spectrum is extracted from the first 30~s of the observation 
  ID.~00569022000 (overlapping with the last part of the decaying BAT lightcurve; see Fig.~\ref{fig:trigger}). The best fit model 
  is obtained with an absorbed blackbody model ($\chi^{2}_{\rm red}/d.o.f.$=1.02/233). The fit gave 
  $N_{\rm H}$=(0.7$\pm$0.1)$\times$10$^{22}$~cm$^{-2}$, $kT$=2.4$\pm$0.1, and $R_{\rm BB}$=7.3$\pm$0.5~$d_{7.2 \rm kpc}$~km. 
  The residuals from the fit are shown in the bottom panel. The estimated 0.5-50~keV X-ray flux is 
  4.5$\times$10$^{-8}$~erg~cm$^{-2}$~s$^{-1}$ (not corrected for absorption).}    
  \label{fig:batxrt}
\end{figure}

We extracted 19 XRT spectra for the first orbit of the XRT observation ID.~00569022000 collected in WT mode 
(total exposure time 1668~s; see Fig.~\ref{fig:trigger}). The integration time of these spectra ranged from a few up to 
300~s, depending on the source flux during the burst decay (see Fig.~\ref{fig:fig_xrtlc_zoom}).  
All 19 spectra could be well fit together by using an absorbed blackbody model ($\chi^{2}_{\rm red}/d.o.f.$=1.10/1863). 
We let the absorption column density free to vary in the fit but tied the value of this parameter to be the same 
for all spectra ($N_{\rm H}$=(0.84$\pm$0.03)$\times$10$^{22}$~cm$^{-2}$). For comparison, we tested that a fit with an absorbed power-law model 
gave an unacceptable result ($\chi^{2}_{\rm red}/d.o.f.$=1.41/1863). The results of this time resolved analysis are summarized in 
Fig.~\ref{fig:fig_xrtlc_zoom}. We also show in Fig.~\ref{fig:batxrt} the quasi simultaneous spectrum obtained from the BAT trigger 
(exposure time 113~s centered around the peak of the event) and the XRT spectrum extracted during the first 60~s of the observation 
ID.~00569022000 (overlapping with the last part of the burst decay as detected by BAT; see Fig.~\ref{fig:trigger}). 
By fitting only the BAT spectrum extracted from the $t=0$ in Fig.~\ref{fig:trigger} until the beginning of the first 
XRT observation (exposure time 170~s), we obtained 
a blackbody temperature of $kT$=2.5$\pm$0.2~keV, a radius of $R_{\rm BB}$=6.7$^{+2.4}_{-1.8}$~$d_{7.2 \rm kpc}$~km, 
and a flux of (5.1$\pm$0.5)$\times$10$^{-9}$~erg~cm$^{-2}$~s$^{-1}$ in the 15-50 keV energy range (corresponding to an extrapolated 
flux of (3.6$\pm$0.4)$\times$10$^{-8}$~erg~cm$^{-2}$~s$^{-1}$ in the 0.5-100 keV energy range). 
The BAT spectrum extracted during the peak of the event (exposure time 1~sec) gave instead 
$kT$=3.0$_{-0.7}^{+1.0}$, $R_{\rm BB}$=6$_{-6}^{+10}$~$d_{7.2 \rm kpc}$~km, and a 15-50~keV 
flux of (1.3$\pm$0.4)$\times$10$^{-8}$~erg~cm$^{-2}$~s$^{-1}$. The latter would correspond to an extrapolated flux of 
(6.0$\pm$1.8)$\times$10$^{-8}$~erg~cm$^{-2}$~s$^{-1}$ in the energy range 0.5-100 keV. 
Based on the spectral analysis and the flux evolution, we suggest in Sect.~\ref{sec:discussion} that 
the event recorded by XRT during the observation ID.~00569022000 is  
a long type-I X-ray burst of a NSLMXB. In Table~\ref{tab:burst} we report the most relevant properties of the burst. 
\begin{table}[htb] 
\caption{Parameters of the long type-I X-ray burst.} 
\label{tab:burst} 
\begin{center} 
\renewcommand{\footnoterule}{} 
\begin{tabular}{ll} 
\hline \hline 
\noalign{\smallskip} 
$F_{\rm peak}^{a}$ (erg cm$^{-2}$ s$^{-1}$) &  (6.0$\pm$1.8)$\times$10$^{-8}$ \\ 
$f_{\rm b}^{b}$ (erg cm$^{-2}$)  &  (1.10$\pm$0.10)$\times$10$^{-5}$\\ 
$\tau_1 \equiv f_{\rm b}/F_{\rm peak}$ (sec)  & 178$_{-53}^{+98}$ \\  
$\gamma  \equiv F^{c}_{\rm pers}/F_{\rm peak}$ & $(8.7\pm4.2)\times10^{-3}$\\ 
\hline 
\end{tabular} 
\end{center} 
\small $^{a}$Unabsorbed flux (0.5-100~keV). 
\small $^{b}$Fluence (0.5-100~keV). 
\small $^{c}$Unabsorbed persistent flux $F_{\rm pers}=(5.2\pm0.2)\times 10^{-10}$ erg cm$^{-2}$ s$^{-1}$ (0.5-100~keV), 
as extrapolated from the closest XRT observation after the burst, i.e. ID.~00569022001.  
\end{table}   

The source X-ray spectra extracted from all the XRT data collected after the first orbit of the observation ID.~00569022000  
could not be well fit by using an absorbed blackbody 
model. In all these cases, an absorbed power-law model provided a better description of the data (see Table~\ref{tab:swift}). 
The power-law spectral index varied along the outburst from 1.6 to $\sim$3, and there is some hint for a  
decrease of the absorption column density in the direction of the source 
from 1.5$\times$10$^{22}$~cm$^{-2}$ to 0.8$\times$10$^{22}$~cm$^{-2}$ during the latest available observations.  
    
We searched all XRT data for possible coherent pulsations and/or quasi-periodic oscillations, but no significant 
feature was found. We obtained the most stringent upper limit on the detection of coherent modulations 
by using the XRT WT data from the observation ID.~00569022000. In this case we estimated a 3$\sigma$ c.l. 
upper limit of 5.6\% on the pulsed fraction of pulsations in frequency range 0.1-281~Hz \citep[the pulsed fraction is 
defined as in][]{vaughan94}. 

\begin{figure}
  \includegraphics[width=6.3cm,angle=-90]{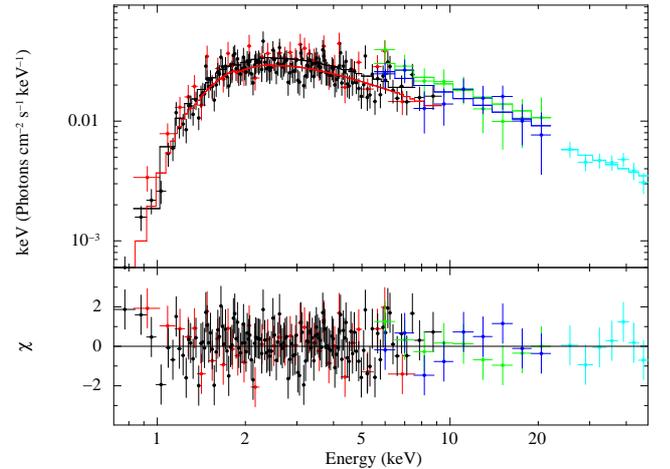}
  \caption{As an example, we show in this figure the broad-band spectrum of \igr\ obtained by using the XRT WT (red) and PC (black) spectra from 
  the observation ID.~00569022006, together with the ISGRI (cyan) and JEM-X (green for JEM-X1 and blue for JEM-X2) 
  data from revolution 1330. The best fit is obtained by using a simple absorbed power-law model. 
  The residuals from the fit are reported in the bottom panel.}  
  \label{fig:broadband}
\end{figure}

We used the quasi-simultaneous \swift\ observation ID.~00569022006 and \inte\ data collected in revolution 1330 to study 
the broad band spectrum of the source at the later stages of the outburst.  
We fit together the XRT spectra in PC and WT mode, with the ISGRI and JEM-X spectra. The best fit  
could be obtained by using a simple absorbed power-law model (see Fig.~\ref{fig:broadband}). 
Normalization constants were included in the fit to take into account the 
inter-calibrations of the instruments and the source variability (data were not strictly simultaneous; see Tables \ref{tab:integral} and 
\ref{tab:swift}). Taking advantage of the other quasi-simultaneous data, we also performed a similar broad-band spectral analysis 
by using: (a) \inte\ data from revolution 1345 and the \swift\ observation ID.~00032930017; (b) \inte\ data from 
revolution 1346 and the \swift\ observation ID.~00032930019; (c) \inte\ data from revolution 1348 and the 
\swift\ observation ID.~00032930021. In all cases a simple absorbed power-law provided a good fit to the data.
All the results are reported in Table~\ref{tab:results_summary}.
\begin{table*}
\caption{Combined \inte\ and \swift\ broad-band spectral analysis of \igr\ (see also Sect.~\ref{sec:XRT} for details). 
$C_{\rm XRT_{\rm PC}}$,  $C_{\rm JEM-X1}$, $C_{\rm JEM-X2}$, and $C_{\rm ISGRI}$ are the normalization constants of the 
different instruments compared to the XRT spectrum (for which the constant was fixed to unity in the fit).} 
\label{tab:results_summary} 
\begin{center} 
\renewcommand{\footnoterule}{} 
\begin{tabular}{llllllllll} 
\hline 
\inte\ rev. & XRT obs. & $N_{\rm H}$ & $\Gamma$ & $C_{\rm XRT_{\rm PC}}$ & $C_{\rm JEM-X1}$ & $C_{\rm JEM-X2}$ & $C_{\rm ISGRI}$ & $F_{\rm 20-50 keV}^{a}$ & $\chi^{2}_{\rm red}/d.o.f.$\\
            &          & (10$^{22}$~cm$^{-2}$) & &                        &                  &                  &                 & & \\
\hline 
\noalign{\smallskip} 
1330       & 00569022006 & 1.30$\pm$0.10 & 1.90$\pm$0.10 & 0.87$\pm$0.06 & 1.4$\pm$0.3 & 1.2$\pm$0.2 & 0.9$\pm$0.2 & 7.7 & 0.87/198\\
\hline 
\noalign{\smallskip}
1345       & 00032930017 & 0.88$\pm$0.06  & 2.26$\pm$0.08  & ---   &  1.0$\pm$0.3  & 1.5$\pm$0.3   &  1.6$\pm$0.4  & 6.0  & 0.96/250\\
\hline 
\noalign{\smallskip}
1346       & 00032930019 & 0.94$\pm$0.05  & 2.34$\pm$0.07  & ---  & ---  & 1.1$\pm$0.2   & 0.9$\pm$0.2   &  6.9 & 0.88/306 \\
\hline 
\noalign{\smallskip}
1348       & 00032930021 & 0.75$\pm$0.05  &  2.26$\pm$0.08 &  --- & 2.2$\pm$0.8  &  3.2$\pm$0.7  &  1.6$\pm$0.4  & 6.1  & 1.0/244 \\
\noalign{\smallskip} 
\hline 
\end{tabular} 
\end{center} 
\small $^{a}$The flux is not corrected for absorption and given in units of 10$^{-10}$~erg~cm$^{-2}$~s$^{-1}$. 
\end{table*}   
All these spectra showed that the broad-band emission from the source was significantly harder during the later 
stages of the event compared to the peak of the outburst observed by BAT (that we interpreted as  
a long type-I X-ray burst; see Sect.~\ref{sec:discussion}). 
To further confirm these findings, we extracted the ISGRI spectrum of the source from rev. 1329. This is the only \inte\ 
detection of the source before the BAT discovery. Although the source was relatively faint in the ISGRI energy band 
during rev. 1329, a sufficiently high S/N spectrum could be extracted by using reduced energy binning (8 channels). A fit 
to this spectrum gave a power-law photon index of 2.3$\pm$0.7 and a 20-50~keV flux of 1.5$\times$10$^{-10}$~erg~cm$^{-2}$~s$^{-1}$.  
 \begin{table*} 	
 \begin{center} 
 \scriptsize
 \caption{Log of all \swift\/XRT observations use in this work. For each observation we also report the results of the spectral fit 
 carried out by using a simple absorbed power-law model (phabs*pow in {\sc XSPEC}).}  	
  \label{tab:swift} 	
 \begin{tabular}{lllllllll} 
 \hline 
 \hline 
 \noalign{\smallskip} 
 Sequence   & Obs/Mode  & Start time  (UT)  & End time   (UT) & Exposure & $N_{\rm H}$ & $\Gamma$ & $F_{\rm 0.5-10\,keV}$$^a$ & $\chi_{\rm red}^2$/d.o.f. \\ 
   &     & (MJD)  & (MJD)  &(s) & (10$^{22}$~cm$^{-2}$) &  & (10$^{-10}$~erg~cm$^{-2}$~s$^{-1}$) &    \\
 \noalign{\smallskip} 
 \hline 
 \noalign{\smallskip} 
00569022000$^b$	&	XRT/WT	&	56536.38578	&	56536.40508	&	1668  \\
00569022000	&	XRT/PC	&	56536.44215	&	56536.58503	&	5581	& 1.49$\pm$0.10 & 2.39$\pm$0.10 & 1.6$\pm$0.6 & 0.92/211$^c$ \\
00569022001	&	XRT/WT	&	56536.58612	&	56536.64248	&	395	& 1.2$\pm$0.2 & 2.3$\pm$0.3 & 1.26$\pm$0.09 & 1.11/145$^c$ \\
00569022001	&	XRT/PC	&	56536.58623	&	56536.67200	&	3698	& 1.5$\pm$0.2 & 2.4$\pm$0.2 & 1.14$\pm$0.06 & 1.11/145$^c$ \\
00569022002	&	XRT/WT	&	56537.17595	&	56537.37689	&	1412	& 1.4$\pm$0.2 & 2.3$\pm$0.2 & 1.11$\pm$0.106 & 0.91/186$^c$ \\
00569022002	&	XRT/PC	&	56537.17604	&	56537.25757	&	2565	& 1.4$\pm$0.2 & 2.1$\pm$0.2 & 1.1$\pm$0.1 & 0.91/186$^c$ \\
00569022003$^d$	&	XRT/WT	&	56537.64302	&	56537.71031	&	153	& --- & --- & --- & ---  \\
00569022003	&	XRT/PC	&	56537.64434	&	56537.72605	&	3814	& 1.24$\pm$0.13 & 1.93$\pm$0.11 & 2.00$\pm$0.07 & 0.97/152 \\
00569022004$^d$	&	XRT/WT	&	56538.11015	&	56538.33109	&	324	& --- & --- & --- & --- \\
00569022004	&	XRT/PC	&	56538.11176	&	56538.33635	&	3638	& 1.01$\pm$0.09 & 1.62$\pm$0.10 & 3.00$\pm$0.10 & 1.05/185 \\
00569022006	&	XRT/WT	&	56539.38610	&	56539.45124	&	159	& 1.06$\pm$0.29 & 1.7$\pm$0.3 & 3.3$\pm$0.3 & 0.91/184$^c$ \\
00569022006	&	XRT/PC	&	56539.38740	&	56539.46243	&	2164	& 1.36$\pm$0.13 & 1.88$\pm$0.13 & 2.81$\pm$0.12 & 0.91/184$^c$ \\
00569022007	&	XRT/WT	&	56540.51553	&	56540.59720	&	1890	& 1.33$\pm$0.14 & 2.25$\pm$0.13 & 1.10$\pm$0.06 & 1.06/162 \\
00569022008	&	XRT/WT	&	56541.26421	&	56541.27637	&	1024	& 1.6$\pm$0.3 & 3.0$\pm$0.3 & 0.34$\pm$0.03 & 1.04/40 \\
00569022009	&	XRT/WT	&	56542.73167	&	56542.74373	&	1021	& 1.4$^{+4.8}_{-1.3}$ & 2.5$^{+3.3}_{-1.5}$ & 0.03$^{+0.01}_{-0.02}$ & 1.32/5 \\
00032930001$^e$ &  XRT/WT	& 56543.66528	&  56543.67847 &  925   & 1.4 & 2.5 & $\sim$0.01 &---  \\
00032930002$^e$ &  XRT/WT	& 56544.60000	&  56544.61319 &  1068  & 1.4 & 2.5 & $\sim$0.005 &---  \\
00032930003$^f$ &  XRT/PC	& 56545.60069	&  56545.61389 &  1096  & 1.6 & 2.5 & $<$0.005 &---  \\
00032930004$^f$ &  XRT/PC	& 56546.32569	&	56546.33889 &  973   & 1.6 & 2.5 & $<$0.006 &---   \\
00032930005$^f$ &  XRT/PC	& 56549.46944	&	56549.88472 &  3814  & 1.6 & 2.5 & $<$0.002 &---   \\
00032930006$^f$ &  XRT/PC	& 56550.39653	&	56550.54861 &  3936  & 1.6 & 2.5 & $<$0.006 &---   \\
00032980001 &  XRT/PC   & 56572.28104  &  56572.30564 &  1973  & 1.6$\pm$0.2 & 2.3$\pm$0.2 & 1.68$\pm$0.07 & 1.14/76  \\
00032980002 &  XRT/PC   & 56573.22292  &  56573.29167 &  1878  & 1.5$\pm$0.2 & 2.4$\pm$0.2 & 1.00$\pm$0.06 & 0.82/45   \\
00032930007 &  XRT/WT	& 56576.83264	&	56576.84583 &  947   & 1.1$\pm$0.2 & 2.2$\pm$0.2 & 1.15$\pm$0.08 & 1.03/84   \\
00032930008 &  XRT/WT	& 56577.63889	&	56577.65208 &  980   & 1.27$\pm$0.14 & 2.10$\pm$0.14 & 1.82$\pm$0.08 & 0.92/127   \\
00032930009 &  XRT/WT	& 56578.55903	&	56578.57222 &  944   & 1.6$\pm$0.4 & 2.4$\pm$0.4 & 0.89$\pm$0.07 & 0.85/31 \\
00032930010 &  XRT/WT   & 56579.50069	&	56579.51458 &  1126  & 1.26$\pm$0.14 & 2.06$\pm$0.12 & 1.87$\pm$0.08 & 1.05/152  \\
00032930011 &  XRT/WT	& 56580.63472	&  56580.64931 &  1044  & 1.2$\pm$0.2 & 2.2$\pm$0.2 & 1.02$\pm$0.06 & 0.89/93 \\
00032930013 &  XRT/WT	& 56582.09583	&	56582.10833 &  913   & 1.3$\pm$0.2 & 2.2$\pm$0.2 & 1.17$\pm$0.07 & 0.98/85\\
00032930014 &  XRT/WT	& 56583.43056	&	56583.44444 &  1064  & 1.15$\pm$0.11 & 2.04$\pm$0.11 & 2.21$\pm$0.07 & 0.88/159 \\
00032930015 &  XRT/WT	& 56584.56458	&	56584.57847 &  1025  & 0.93$\pm$0.09 & 2.00$\pm$0.10 & 2.40$\pm$0.07 & 1.08/171  \\
00032930016 &  XRT/WT	& 56585.97014	&	56585.98333 &  961   & 0.91$\pm$0.08 & 2.28$\pm$0.10 & 2.80$\pm$0.10 & 0.86/176 \\
00032930017 &  XRT/WT	& 56586.04097	&	56586.05417 &  960   & 0.88$\pm$0.06 & 2.27$\pm$0.08 & 3.23$\pm$0.11 & 0.98/222 \\
00032930018 &  XRT/WT	& 56587.44236  &	56587.45556 &  526   & 0.74$\pm$0.07 & 2.14$\pm$0.11 & 3.2$^{+0.1}_{-0.2}$ & 0.99/142 \\
00032930019 &  XRT/WT	& 56588.37708	&	56588.51667 &  1049  & 0.94$\pm$0.05 & 2.36$\pm$0.07 & 4.63$\pm$0.10 & 0.89/292  \\
00032930021 &  XRT/WT	& 56594.45208	&	56594.46458 &  946   & 0.76$\pm$0.05 & 2.27$\pm$0.08 & 3.00$\pm$0.08 & 1.06/221 \\
00033239001$^f$ &  XRT/PC   & 56763.28194  &  56763.42292 &  2021  & 0.76 & 2.27 & $<$0.004 & ---   \\
\hline
\hline
  \end{tabular}
  \end{center}
  \begin{list}{}{} 
  \scriptsize
  \item[$^{\mathrm{a}}$:] Observed flux (i.e., not corrected for absorption); 
  \item[$^{\mathrm{b}}$:] Analyzed separately (see Sect.~\ref{sec:swift} for details);
  \item[$^{\mathrm{c}}$:] We fit the PC and WT spectrum together, leaving all parameters free to vary but obtaining a single 
  $\chi_{\rm red}^2/d.o.f.$ from the fit; 
  \item[$^{\mathrm{d}}$:] Statistics was too poor to extract a meaningful spectrum; 
  \item[$^{\mathrm{e}}$:] The source was detected but too faint for the XRT WT mode. A rough estimate of the source flux is given from the 
  recorded count rate and assuming the spectral model of the previous observation with good statistics. 
  \item[$^{\mathrm{f}}$:] The source was not detected and thus we used the spectral model of the preceding observation to derive 
  a 3$\sigma$ upper limit on the source X-ray flux.
  \end{list} 
  \end{table*}

\begin{figure}
\includegraphics[width=4.6cm,angle=-90]{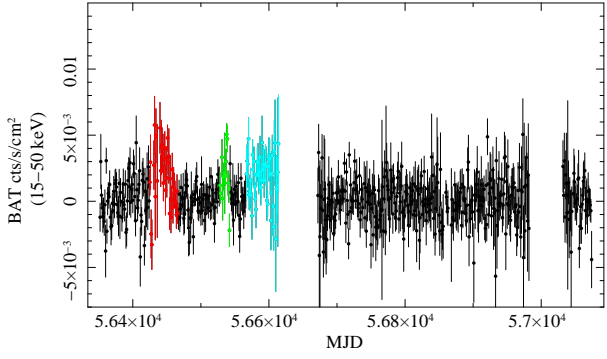}
\caption{The long term BAT lightcurve of \igr\ in the 15-50~keV energy range. 
The two outbursts of the source in May-June 2013 (red points) and September 2013 (green and cyan points) are 
clearly visible. For the outburst in September, we highlighted in green the initial part of the event up to the 
quiescent period around 56545~MJD. The cyan points mark the re-brightening of the source after the quiescent 
period up to 56614~MJD. Gaps in the BAT lightcurve are due to the source proximity to the Sun.}   
\label{fig:bat}
\end{figure}

For completeness, we also reported in Fig.~\ref{fig:bat} the long term BAT lightcurve of the source as retrieved from the 
BAT Transient Monitor web-page\footnote{\href{http://swift.gsfc.nasa.gov/results/transients/weak/SwiftJ1734.5-3027/}
{http://swift.gsfc.nasa.gov/results/transients/weak/SwiftJ1734.5-3027/}.}   
\citep{krimm13}. The lightcurve starts on MJD 56352 (i.e. 2013 March 1). The outburst reported in this paper 
and the previous detection of \igr\ in May-June 2013 are clearly visible. The latter detection was also discussed by \citet{laparola13}. 
These authors confirmed that the BAT spectral index at that time was consistent with that measured by ISGRI before the onset 
of the bright event recorded in September 2013 and during the latest stages of such event 
(see Sect.~\ref{sec:intro} and \ref{sec:XRT}). Unfortunately, no pointed XRT observations were performed in May-June 2013.  

\subsection{UVOT}
\label{sec:uvot}

We also analyzed all available UVOT data that were collected simultaneously with the XRT observations. 
 The analysis of the UVOT data was performed using the {\sc uvotimsum} and
{\sc uvotsource} tasks included in the {\sc FTOOLS} software.  
The latter task performs aperture photometry on a single source 
based on user-supplied source and background 
regions\footnote{\href{http://heasarc.nasa.gov/ftools/caldb/help/uvotsource.html}
{http://heasarc.nasa.gov/ftools/caldb/help/uvotsource.html}}; 
we adopted circular regions of 5\arcsec{} and 10\arcsec{} for source and 
background, respectively. 
A faint source is found within the XRT error circle in the coadded UVOT $u$ 
images (total exposure 5977\,s).  For this source we obtained a magnitude of 
$u=19.70\pm 0.12$ mag. 
The best estimated UVOT position for this object is at
RA(J$2000)$ = $17^{\rm h} 34^{\rm m} 24\fs33$, 
Dec(J$2000)=-30^{\circ} 23^{\prime} 53\farcs2$   
with an uncertainty of 0\farcs43 (90\,\% c.l.).  
This is compatible at 2.4~$\sigma$ c.l. with the XRT position reported in 
Sect.~\ref{sec:XRT} \citep[assuming the XRT uncertainty in the position follows a Rayleigh distribution;][]{evans14}.   
Only poorly constraining upper limits could be obtained on the source 
magnitudes in the other UVOT filters. 
Due to the lack of any detection of changes in magnitude of the UVOT source during the outburst 
and the marginal compatibility between the UVOT and XRT position, 
we are currently unable to firmly associate the counterpart identified in this section with 
\igr.\

\section{\xmm\ data}
\label{sec:xmm}

\begin{figure}
  \includegraphics[width=5.8cm,angle=-90]{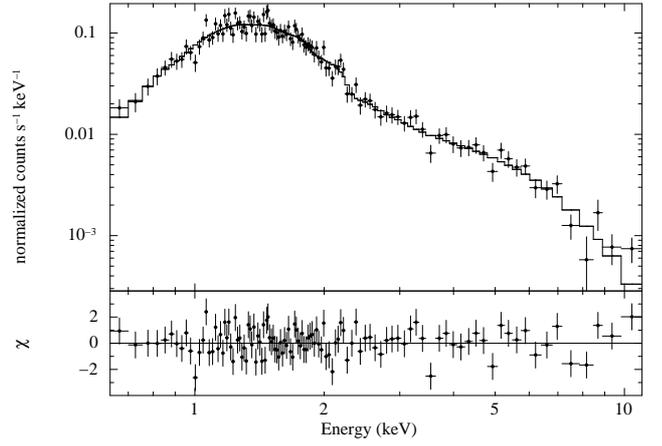}
  \caption{\xmm\ spectrum of \igr\ obtained by summing up the Epic-pn, MOS1, and MOS2 data. The best fit model is obtained by using 
  an absorbed power law plus a blackbody. The residuals from the fit are shown in the bottom panel.} 
  \label{fig:spectra}
\end{figure}

An \xmm\ \citep{jansen01} observation of \igr\ was carried out on 2013 September 8 for a total exposure time of 32.3~ks 
(from 56543.1404~MJD to 56543.5293~MJD).  
The EPIC-pn was operated in timing mode, the MOS1 in full frame and the MOS2 in small window. 
We reduced these data by using the SAS version 13.5 and the latest \xmm\ calibrations files 
available\footnote{\href{http://xmm2.esac.esa.int/external/xmm_sw_cal/calib/index.shtml}
{http://xmm2.esac.esa.int/external/ xmm\_sw\_cal/calib/index.shtml}}. The observation was not affected by intervals of 
high flaring background; we thus retained the entire exposure time available for the following analysis. As the source was found to be 
relatively faint (see later in this section) and centered on {\sc RAWX=37}, the EPIC-pn  
lightcurve and spectra were extracted in the energy range 0.6-12~keV from CCD columns 36 through 38 for the source and 3 through 15 for the 
background. 
For the two MOS cameras we used a circular source (background) extraction region with a radius of 600 (1800) pixels, corresponding 
to $30''$ ($90''$). The average count-rate of the source as recorded by the EPIC-pn was (8.3$\pm$0.2)$\times$10$^{-2}$ cts~s$^{-1}$, thus 
too low to perform any count-rate and/or time resolved analysis. To improve the statistics, we used the task {\sc epicspeccombine} to combine 
all EPIC spectra and fit them together (see Fig.~\ref{fig:spectra}).  
A fit to this spectrum with a simple absorbed power-law model gave an unacceptable result with $\chi_{\rm red}^2$/d.o.f.=1.88/111. 
The addition of a thermal blackbody component ({\sc bbodyrad} in {\sc xspec}) significantly improved the fit 
($\chi_{\rm red}^2$/d.o.f.=1.07/109). We measured an absorption column density of $N_{\rm H}$=(0.57$\pm$0.08)$\times$10$^{22}$ cm$^{-2}$, 
a power-law photon index of $\Gamma$=1.2$\pm$0.3, a blackbody temperature of $kT$=0.32$\pm$0.03~keV, and a radius 
of $R_{\rm BB}$=(1.5$^{+0.5}_{-0.3}$)$\times$$d_{7.2 \rm kpc}$~km.   
The average absorbed (unabsorbed) 0.5-10~keV X-ray flux measured from the spectral fit is 
5.5$\times$10$^{-13}$~erg~cm$^{-2}$~s$^{-1}$ (8.9$\times$10$^{-13}$~erg~cm$^{-2}$~s$^{-1}$). This is 
consistent with the approximate measurement of the source flux obtained with XRT in the observations ID.~00032930001 and 
00032930002 (see Table~\ref{tab:swift}). The $N_{\rm H}$ measured by \xmm\ suggests that the absorption 
column density in the direction of the source might slightly decrease not only during the latest stages of the outburst, but also 
during the low X-ray luminosity period (unfortunately during this period we could not obtain a reliable $N_{\rm H}$ measurement 
with XRT due to the low statistics of the data; see Sect.~\ref{sec:swift}).  

We searched for possible coherent pulsations and/or quasi-periodic oscillations in the \xmm\ data, 
but none was found (also due to the very limited statistics). For coherent modulations, we estimated  
an upper limit of 13.5\% at 3$\sigma$ c.l. on the pulsed fraction of pulsations in the frequency range 
0.1-1500\,Hz. 

For completeness, we checked for possible evidence of the soft thermal component detected by \xmm\ in the XRT data collected close  
to 56543~MJD. In the XRT observations ID 00032930001-00032930006 the source was not detected. 
We thus considered the observation ID.\,00569022007, which was performed 2.4 days earlier than the \xmm\ 
one and was characterized by a reasonably high statistics (see Table~\ref{tab:swift}). 
The addition of a blackbody component to the source spectrum in this observation did not improve the fit. 
We thus fixed the temperature of this component to the one measured by \xmm\ and determined a 90\% c.l. upper limit on its  
normalization. We obtained $R_{\rm BB}$$<$7~km (for an assumed distance of 7.2~kpc). This result suggests that the thermal 
emission revealed by \xmm\ might have gone undetected during the earlier stages of the outburst due to the 
higher flux of the power-law component. We comment more on this result in Sect.~\ref{sec:discussion}.

\section{Discussion}
\label{sec:discussion}

In this paper we report on all X-ray data collected during the first followed-up outburst of the source \igr,\ which also led to its 
discovery. Shortly after the onset of the outburst, the source displayed a long type-I X-ray burst that we discuss in detail  
later in this section. This allows us to firmly classify the source as a new member of the bursting NSLMXBs. 

As noticed in Sect.~\ref{sec:swift}, the source showed a remarkable decay in flux after the onset of the outburst up to 56539~MJD. 
At this epoch, the decay of the flux steepened dramatically and the source faded below the XRT detection limit 
on $\sim$56543~MJD (see Fig.~\ref{fig:fig_xrtlc}). 
Such outburst profile is, in several respects, qualitatively similar to those observed from transient NSLMXBs. The 
outbursts of these systems typically last for a few weeks and are characterized by an initial exponential decay. 
At the later stages, a break is sometimes observed in the flux decay and a knee in the lightcurve marks the transition to a faster 
linear decaying phase that lasts until the source fades back into quiescence\footnote{In some cases, a linear decay in flux is 
observed directly after a stage in which the flux of the source is relatively stable for several weeks 
\citep[see, e.g.,][]{heinke15,campana13}.} \citep[see e.g.,][]{gilfanov98,gierlienski05,falanga05, 
falanga05b,falanga11,falanga12}. These outburst profiles are usually interpreted in terms of the 
disk instability model proposed by \citet{king98}. These authors showed that the X-ray heating 
during the decay from an outburst causes the light curves of a transient LMXBs to exhibit
an exponential decay if the X-ray luminosity is sufficiently high to maintain the outer disk radius hot, or a linear 
decay at lower luminosities. This model was successfully applied to a number of observed LMXB outbursts 
by \citet{powell07}, who also showed that the luminosity at which the knee occurs in the lightcurve  
and the exponential timescale of the outburst decay can be used to estimate the source outer disk radius.  
Indeed, it is believed that the matter supply to the NS is effectively 
cut-off after the knee, when the irradiation of the outer disk radius diminishes and the disk 
enters a cool low-viscosity state. 
 
The peculiarity of the \igr\ outburst profile is the presence of an apparently quiescent period extending from 
56545.6~MJD to at least 56550.5~MJD that preceded a new rebrightening. During the quiescent period the source 
remained in a flux level that was too low to be detectable within the short XRT exposures (see Table~\ref{tab:swift}). 
Stacking together the XRT observations ID.~00032930003, 00032930004, 
00032930005, and 00032930006 a 3$\sigma$ upper limit on the source count rate of 0.0049 cts~s$^{-1}$ is obtained, 
corresponding to an X-ray luminosity of 1.6$\times$10$^{33}$~erg~s$^{-1}$ (at 7.2~kpc, see below).
As LMXBs in quiescence might have a luminosity as low as $\lesssim$10$^{32}$~erg~s$^{-1}$, it 
is not possible to determine if \igr\ really faded back into quiescence during this period or underwent an episode 
of very low mass accretion rate. 
Surprisingly, the source resumed its X-ray active state around 56572~MJD. There was no coverage available with XRT 
for the period 56550-56572~MJD, but the \inte\ data show that the high intensity state of the source did not 
resume before 56565~MJD. The apparent low activity period in X-rays thus lasted about 15~days. 
At the beginning of the re-brightening phase the source recovered roughly the same X-ray luminosity displayed immediately 
after the long type-I X-ray burst, and its broad-band X-ray spectrum was found to be fully compatible with that observed during most of 
the previous outburst. The mechanism that triggered this second episode of enhanced accretion is difficult to investigate due to the paucity 
of the data and the lack of coverage during the rise and the decay from this renewed activity. The last available XRT observation of \igr\ was 
carried out on 56763~MJD, i.e. about 169~days after the last detection of the source in the X-ray active phase, and thus we do not 
know precisely how long the second rebrightening phase lasted in time.

We note that a similar ``double'' outburst as described above was observed from the accreting millisecond X-ray pulsar (AMXP) IGR\,J00291+5934, 
which underwent two $\sim$10~days long outbursts in 2008 separated by a quiescent period of $\sim$30~days  \citep{hartman11}. 
In this case, the first outburst followed the usual profile displaying a knee during the decay at a certain luminosity, 
while the second was characterized by a significantly slower rise and decay times. Due to the presence of the second outburst, 
\citet{hartman11} questioned the applicability of the irradiated disk model to IGR\,J00291+5934. The main issue with this 
interpretation is that it predicts the disk around the NS to be heavily emptied after the first outburst and unable to  
power a second event without invoking an unlikely large variation in the mass flow rate from the companion star. 
The authors suggested that the knee during the decay of the first outburst could have been due to the onset of a propeller 
stage, during which accretion is (at least) partly inhibited due to the rapidly rotating NS 
magnetosphere \citep[see, e.g,][and references therein]{illarionov75, romanova14}. The inhibition of accretion 
would leave enough material in the disk after the first outburst to fuel a second event within a few days. 
As the condition for the onset of a propeller effect depends strongly on the NS magnetic field strength and rotational 
period, it is difficult to test this scenario in the case of \igr,\ for which neither the spin period nor its derivative 
are known \citep[the NS magnetic field can be inferred from measurements of the spin period variations over 
time; see, e.g,][and references therein]{papitto07}. Furthermore, it is worth noticing that, despite our understanding of the 
propeller regime is still far from be complete, this effect is often involved to interpret more abrupt drops in luminosity 
\citep[see, e.g.,][for the recent case of the AMXP IGR J18245-2452]{ferrigno14}.    

The \xmm\ spectrum of the source extracted during the faster portion of the first outburst decay (around 56543~MJD) could 
not be well described by a simple absorbed power-all model, at odds with all previously collected XRT spectra.  
The \xmm\ data revealed the presence of a predominant thermal component, which properties are compatible with those 
expected for an emitting hot spot on the NS surface (see Sect.~\ref{sec:xmm}). 
As discussed in Sect.~\ref{sec:xmm}, we could not firmly rule out the presence of this thermal component in the 
XRT data; if present, the thermal emission could have been likely hidden by the dominant  
comptonization component (i.e. the power-law due to the accretion).  
We thus conclude that during the latest stages of the first outburst, accretion was strongly suppressed  
and the emission from the NS surface emerged at the soft X-rays ($\lesssim$2~keV). We are, however, currently unable  
to discriminate between a suppression of the accretion due to the disk turning into the low viscosity state 
or the onset of a mild propeller phase. 

It is interesting to note that, at odds with a number of previously studied systems \citep{kuulkers09}, 
the long type-I burst burst recorded by \igr\ is unlikely to be the event triggering the overall outburst of the 
source. The \inte\ observations carried out in rev. 1329 already detected an enhanced X-ray activity from the source 
a few days before the onset of the long type-I burst. This indicates that the NS was already 
accreting a substantial amount of matter at the time. We thus conclude that   
the outburst developed as a consequence of the disk instability, and it was the latter that 
paved the way for the long type-I X-ray burst rather than the other way around.  
As bursts from \igr\ were never detected before of the outburst in September 2013, 
it is unlikely that the source undergoes thermonuclear runaways during its quiescent 
state, as observed from the so-called ``burst-only sources'' \citep[see, e.g.,][]{cocchi01}. 
This would be expected according to the known stability 
of hydrogen burning via the pp-process or pycnonuclear reactions at low X-ray luminosities \citep{fushiki87}. 
Assuming a quiescent luminosity for \igr\ of $\sim$10$^{33}$~\ergsec (as estimated from the XRT observations and assuming 
a distance of $\sim$7.2~kpc; see Sect.~\ref{sec:distance}), it would take $\sim$200 years to the source 
to accumulate the necessary fuel on the neutron star surface to emit a 
burst with a characteristic energy release of $\sim$6.7$\times$10$^{40}$~erg (see Sect.~\ref{sec:distance}). 
Therefore, it seems that \igr\ gives rise to thermonuclear explosions only when it is accreting at luminosities 
$\gtrsim$10$^{35}$-10$^{36}$~\ergsec and is thus a new member of the faint transient NSLMXBs exhibiting pure 
helium runaways when in outburst.

\subsection{The long type-I X-ray burst}
\label{sec:burst}

From the time resolved spectroscopy and flux evolution of the \swift\ data collected on 2013 September 1, we argued that 
the event triggering the BAT on that day was due to a long type-I X-ray burst from \igr.\ The time and spectral 
properties of the event are strikingly similar to those of the long burst from the NSLMXB XTE\,J1701-407 that triggered 
the BAT on 2008 July 17 \citep{falanga09}.   
We discuss the physical properties of the long burst from \igr\ in the next three sub-sections and use our observational findings 
to derive constraints on the distance to the source and the details of the accretion process.

\subsubsection{light curve and spectra}

Our analysis of the XRT and BAT data presented in this paper revealed a long 
type-I X-ray burst from \igr.\ This is the first discovered burst from the source.  
In Fig.~\ref{fig:fig_xrtlc_zoom}, we showed the BAT (upper panel) 
and XRT (lower panel) light curve of the long type-I X-ray burst. 
We set the burst start time at 56536.38458~MJD,  i.e. the time at which
the BAT X-ray intensity of the source increased by 10\% compared to its   
persistent intensity level. The XRT started observing the source about 
170~s after the beginning of the burst detected by BAT. The
BAT light curve exhibits a slow rise time\footnote{The rise time is
  defined as the time spent between the 
start of the burst and the point at which the 90\% of the peak burst
intensity is reached.} of $\approx50$~s.
The XRT decay time from the burst is best-fit by using two exponential functions 
with e-folding times of $\tau_1=110\pm8$~s and $\tau_2=1019\pm85$~s,
respectively. A single exponential function could not provide an acceptable fit to the data. 
We note that the value  $\tau_1 \equiv f_{\rm b}/F_{\rm peak}=178_{-53}^{98}$~s reported in 
Table~\ref{tab:burst} is roughly consistent with the e-folding time estimated from the fit to the burst profile.  
Similar long type-I X-ray bursts were so far recorded from the sources 2S\,0918-549, SLX\,1737-282, and 
XTE\,J1701-407 \citep{zand05,falanga08,linares08,falanga09}. To compare the burst from \igr\ 
with these previous results and with the decay cooling model proposed by 
\citet{cumming04}, we fitted the burst flux decay profile in Fig.~\ref{fig:fig_xrtlc_zoom} with a power-law, and found an
index of $\Gamma = -1.77\pm0.02$ with a $\chi^{2}_{\rm red}/d.o.f.=1.5/17$.  
This is comparable to the results found in the other cases. 
The total duration of the burst, i.e. the time to evolve away from and
return to the persistent state, was of $\approx200$~s and
$\approx30$~minutes in the BAT (15-25 keV) and the XRT (0.3-10 keV) lightcurves, respectively.  

It is well known that X-ray bursts are produced by unstable burning of
matter accreted on the surface of NSs and the X-ray emission released in this process 
can be described well as blackbody radiation with temperatures of a few keV. 
The energy-dependent decay time of the burst is regulated by the cooling of the NS
photosphere, which results in a gradual softening of the burst spectrum 
\citep[see, e.g.,][]{lewin93,strohmayer06}. The time-resolved spectral analysis
of the burst recorded from \igr\ was carried out by using BAT and XRT data in the
15-25~keV and 0.3--10~keV bands, respectively. Burst spectra
could be described well by a simple photoelectrically-absorbed blackbody model 
(we measured $N_{\rm H}$=0.84$\times$10$^{22}$~cm$^{-2}$; see Table~\ref{tab:burst}).
The inferred blackbody temperature, apparent radius at 7.2~kpc (see below), and
bolometric luminosity are reported in Fig.~\ref{fig:fig_xrtlc_zoom} and in Table~\ref{tab:burst}.    
The burst fluence has been calculated from the bolometric 
fluxes (these correspond to the observed 0.3-10~keV XRT fluxes 
extrapolated to the 0.5-100~keV energy range). 
The peak flux of the burst was derived from the BAT 
spectrum extrapolated to the 0.5-100~keV energy range. 
All measured fluxes during the burst were extrapolated to the
0.5-100~keV band by generating dummy responses (XSPEC v12.8.2). 
This is justifiable for the XRT data since the
blackbody temperature is well inside the spectral energy coverage. 
During the main peak of the burst, the blackbody temperature reaches a maximum  
in the energy range 2-3 keV \citep[see e.g.,][and references therein]{falanga08}, and
thus the bolometric flux of the blackbody emission from the source is outside the BAT bandpass.
However, as we detail below, this introduces uncertainties that are well within the estimated 
error on the bolometric flux measurements reported in Table~\ref{tab:burst}.  
The best-fit blackbody temperature measured from the BAT spectrum extracted during the peak of the burst 
was $kT=3.0_{-0.7}^{+1.0}$~keV. We fixed an upper and lower boundary blackbody temperature of 2.7 and 3.5~keV, 
respectively, the normalization remaining as a free parameter. 
The lower boundary for the temperature is chosen to be higher than the temperature measured during the 
XRT data collected close to the peak of the burst (see Fig.~\ref{fig:fig_xrtlc_zoom}). 
For the upper boundary, we assumed that kT$\sim$3.5~keV is a reasonably high value for the black-body 
temperature at the peak of the event based on previous observations of long bursts from 
LMXBs \citep[see, e.g.,][]{molkov05,chenevez06,chenevez07,falanga08,falanga09}. 
In the two cases, we found extrapolated fluxes of 4.4$\times$10$^{-8}$~\ferg and 7.0$\times$10$^{-8}$~\ferg. 
These are well within the boundary of the peak flux uncertainties reported in Table~\ref{tab:burst}. 
Note that a similar procedure was applied to extrapolate the peak flux of the burst in the case of 
XTE\,J1701-407 \citep{falanga09}.

\subsubsection{Source distance, persistent flux, and accretion rate}
\label{sec:distance} 

In photospheric-radius expansion (PRE) bursts, the source
distance can be determined based on the assumption that the bolometric
peak luminosity is saturated at the Eddington limit  
\citep[$L_{\rm Edd}$; e.g.,][]{lewin93,kuulkers03}. During the PRE episode, the
bolometric luminosity remains virtually constant at the Eddington 
value, while the high energy flux displays a typical double-peak profile 
and/or a delay in the rise time \citep[e.g.,][]{kuulkers02,galloway06,falanga07,falanga08}.

In the case of \igr,\ the BAT lightcurve shows a slow rise time of $\approx50$~s, which is
typically observed at the hard X-rays long type-I helium bursts with PRE
\citep[e.g.,][]{kuulkers02,molkov05,falanga08}. However, the XRT observation started 
about 170 seconds after the peak of the event, and thus we cannot firmly 
assess if this burst underwent a PRE. The time-resolved spectral analysis of the BAT data 
could not provide sufficient indications to solve this issue due to the limited 
statistics (see Fig. \ref{fig:trigger}). From the comparison between the burst from \igr\ with other PRE bursts 
exhibiting a similar slow rise time at high energies, we assume in the following that \igr\ underwent  
a PRE during the first $\approx 200$~s of the event.  
This should correspond to the timescale that is needed by the photosphere to fall back onto the 
NS surface \citep[see discussion in][]{falanga09}.  
Under these assumptions and considering a bolometric peak luminosity equal to 
the Eddington value for a He type-I X-ray burst 
\citep[$L_{\rm Edd}\approx\,3.8\times 10^{38}$ ergs$^{-1}$, as empirically derived by][]{kuulkers03},  
we obtain a source distance for \igr\ of $d=7.2\pm1.5$~kpc. 
We note that the source would be closer if the peak luminosity of the
burst were lower than the pure-helium Eddington 
limit, or if a solar composition ($X_0=0.7$) is assumed for the accreted material 
(in this case the source distance would be $\approx6$~kpc). In the
following, we consider $d\approx7.2$~kpc as a fiducial distance.

In the case of \igr,\ an estimate of the persistent emission immediately before the onset of the long type-I 
burst was not available (see Sect.~\ref{sec:swift}). We thus assume as the pre-burst persistent emission, the flux measured 
in the closest XRT observation (i.e. the PC data in observation ID.~00569022000). 
In these data, the X-ray emission from \igr\ was found to be well described by a simple power-law model 
with $\Gamma \approx2.4$. Assuming a distance of 7.2~kpc, the estimated long type-I burst  
persistent unabsorbed flux in the energy range 0.5-100~keV is thus $F_{\rm pers}\approx~5.2\times 10^{-10}$ erg cm$^{-2}$ s$^{-1}$, 
translating into a bolometric luminosity of $L_{\rm pers}\approx~3.3\times10^{36}$~\ferg, i.e. $\approx1.0\% L_{\rm Edd}$. 
We can estimate the local accretion rate per unit area as $L_{\rm pers}=5.4\pi R^2\dot m(GM/R)/(1+z)$, or 
\begin{eqnarray}
\dot m & \sim & 2\times 10^3\ {\rm g\ cm^{-2}\ s^{-1}}  \nonumber\\
& & \left({R\over 11.2\ {\rm
    km}}\right)^{-1}\left({M\over
  1.4\ M_\odot}\right)^{-1}\left({d\over 7.2\ {\rm
    kpc}}\right)^2\left({1+z\over 1.26}\right). 
\end{eqnarray}
A convenient unit of accretion rate is the Eddington 
rate. Here, we define the local Eddington accretion rate to be  
$\dot m_{\rm Edd}\equiv 1.8\times 10^5\ {\rm g\ cm^{-2}\ s^{-1}}$, 
which corresponds to the local accretion rate of a NS of mass 
$M=1.4\ M_\odot$ and radius $R=11.2\ {\rm km}$ that has a luminosity equal to the   
Eddington value $3.8\times 10^{38}\ {\rm erg\ s^{-1}}$  
\citep{kuulkers03}. We thus conclude that $\dot m/\dot m_{\rm Edd}\sim1$\%.

\subsubsection{The energy, ignition depth, and recurrence time of the burst} 
\label{sec:3-2}
 
The released energy measured during the long-burst allows us to estimate the
ignition depth and derive predictions on the burst recurrence time. The measured fluence of the burst, $f_b=1.1\times 10^{-5}\ {\rm erg\ cm^{-2}}$, 
corresponds to a net burst energy release of $E_{\rm burst}=4\pi d^2f_b=6.7\times 10^{40}\ {\rm ergs}\ (d/7.2\ {\rm kpc})^2$. 
The ignition depth is related to the energy release during the burst by the equation 
$E_{\rm burst}=4\pi R^2y_{\rm ign}Q_{\rm nuc}/(1+z)$, and thus  
\begin{eqnarray}
y_{\rm ign} &=& 3.4\times 10^{9}\ {\rm g\ cm^{-2}} \left({d\over 7.2\ {\rm kpc}}\right)^2\nonumber\\
& &  \left(\frac{Q_{\rm nuc}}{1.6\ {\rm MeV/nucleon}}\right)^{-1}\left(\frac{R}{11.2\ {\rm
    km}}\right)^{-2}\left(\frac{1+z}{1.26}\right).  
\end{eqnarray}
The value of $Q_{\rm nuc}\approx 1.6$ MeV/nucleon corresponds to the nuclear-energy 
release per nucleon for complete burning of helium to iron
group elements. Including hydrogen with a mass-weighted mean mass
fraction $\langle X\rangle$ provides a value of $Q_{\rm nuc}\approx 1.6+4\langle
X\rangle$ MeV/nucleon \citep{galloway04}, where we included losses due to
neutrino emission following \citet{fujimoto87}. For $\langle
X\rangle=0.7$, the solar composition value, $Q_{\rm nuc}=4.4$ MeV/nucleon, and
$y_{\rm ign}=1.2\times 10^9\ {\rm g\ cm^{-2}}$. 
At an accretion rate of $2\times 10^3\ {\rm g\ cm^{-2}\ s^{-1}}$, the
recurrence time corresponding to a column depth of $y_{\rm
  ign}=3.4\times 10^{9}\ {\rm g\ cm^{-2}}$ (pure helium composition) is $\Delta t=(y_{\rm
  ign}/\dot m)(1+z)\sim24\ {\rm days}$, or for $y_{\rm
  ign}=1.2\times 10^{9}\ {\rm g\ cm^{-2}}$ (solar composition) is
$\Delta t\sim9$ days (independent from the assumed distance). 

The long type-I burst, which was the first
observed burst from \igr,\ occurred about half a day after the first 
\inte\ detection of the source in rev. 1329 (starting from 56535.85950~MJD). The outburst of the source 
was closely monitored only for the first 15 days, when the source first faded into a 
low luminosity state (see Sect.~\ref{sec:swift} and Fig.~\ref{fig:fig_xrtlc}). 
The re-brightening of the source was observed starting from 56565.5999~MJD, thus leaving 
a gap in the outburst coverage of about 15~days during which we do not have information on the 
X-ray emission from the source. The last useful XRT observation of the source during the re-brightening 
was carried out up to 56594.46458~MJD, i.e. about 58~days from the detected long type-I burst. 
According to our estimates above, the case $Q_{\rm nuc}\approx 1.6$ MeV/nucleon would be slightly favoured by the data (as we did not 
observe a second type-I X-ray burst), but the coverage of the outburst was far too sparse and limited in time 
to draw any firm conclusion.

\section*{Acknowledgments}

EB thanks the New York University at Abu Dhabi for the kind hospitality during part of this work and Milvia Capalbi 
for her precious support during the analysis of some XRT observations. We thank Phil Evans for useful discussions. 
AP is supported by a Juan de la Cierva fellowship, and acknowledges grants AYA2012-39303, SGR2009-811, and iLINK2011-0303. 
PR acknowledges contract ASI-INAF I/004/11/0. 
We thank N. Schartel and the \xmm\ team for having promptly performed the ToO observation analyzed in this paper.  
We are indebted to the Swift PI and operations team for the continuous support during the monitoring campaign 
of X-ray transients.  

\bibliographystyle{aa}
\bibliography{J1734}

\begin{thebibliography}{49}
\expandafter\ifx\csname natexlab\endcsname\relax\def\natexlab#1{#1}\fi

\bibitem[{{Campana} {et~al.}(2013){Campana}, {Coti Zelati}, \&
  {D'Avanzo}}]{campana13}
{Campana}, S., {Coti Zelati}, F., \& {D'Avanzo}, P. 2013, \mnras, 432, 1695

\bibitem[{{Chenevez} {et~al.}(2006){Chenevez}, {Falanga}, {Brandt},
  {Farinelli}, {Frontera}, {Goldwurm}, {in't Zand}, {Kuulkers}, \&
  {Lund}}]{chenevez06}
{Chenevez}, J., {Falanga}, M., {Brandt}, S., {et~al.} 2006, \aap, 449, L5

\bibitem[{{Chenevez} {et~al.}(2007){Chenevez}, {Falanga}, {Kuulkers}, {Walter},
  {Bildsten}, {Brandt}, {Lund}, {Oosterbroek}, \& {Zurita Heras}}]{chenevez07}
{Chenevez}, J., {Falanga}, M., {Kuulkers}, E., {et~al.} 2007, \aap, 469, L27

\bibitem[{{Cocchi} {et~al.}(2001){Cocchi}, {Bazzano}, {Natalucci}, {Ubertini},
  {Heise}, {Kuulkers}, {Cornelisse}, \& {in't Zand}}]{cocchi01}
{Cocchi}, M., {Bazzano}, A., {Natalucci}, L., {et~al.} 2001, \aap, 378, L37

\bibitem[{{Courvoisier} {et~al.}(2003){Courvoisier}, {Walter}, {Beckmann},
  {Dean}, {Dubath}, {Hudec}, {Kretschmar}, {Mereghetti}, {Montmerle},
  {Mowlavi}, {Paltani}, {Preite Martinez}, {Produit}, {Staubert}, {Strong},
  {Swings}, {Westergaard}, {White}, {Winkler}, \& {Zdziarski}}]{courvoisier03}
{Courvoisier}, T., {Walter}, R., {Beckmann}, V., {et~al.} 2003, \aap, 411, L53

\bibitem[{{Cumming} \& {Macbeth}(2004)}]{cumming04}
{Cumming}, A. \& {Macbeth}, J. 2004, \apjl, 603, L37

\bibitem[{{Evans} {et~al.}(2009){Evans}, {Beardmore}, {Page}, {Osborne},
  {O'Brien}, {Willingale}, {Starling}, {Burrows}, {Godet}, {Vetere}, {Racusin},
  {Goad}, {Wiersema}, {Angelini}, {Capalbi}, {Chincarini}, {Gehrels}, {Kennea},
  {Margutti}, {Morris}, {Mountford}, {Pagani}, {Perri}, {Romano}, \&
  {Tanvir}}]{evans09}
{Evans}, P.~A., {Beardmore}, A.~P., {Page}, K.~L., {et~al.} 2009, \mnras, 397,
  1177

\bibitem[{{Evans} {et~al.}(2014){Evans}, {Osborne}, {Beardmore}, {Page},
  {Willingale}, {Mountford}, {Pagani}, {Burrows}, {Kennea}, {Perri},
  {Tagliaferri}, \& {Gehrels}}]{evans14}
{Evans}, P.~A., {Osborne}, J.~P., {Beardmore}, A.~P., {et~al.} 2014, \apjs,
  210, 8

\bibitem[{{Falanga} {et~al.}(2005{\natexlab{a}}){Falanga}, {Bonnet-Bidaud},
  {Poutanen}, {Farinelli}, {Martocchia}, {Goldoni}, {Qu}, {Kuiper}, \&
  {Goldwurm}}]{falanga05b}
{Falanga}, M., {Bonnet-Bidaud}, J.~M., {Poutanen}, J., {et~al.}
  2005{\natexlab{a}}, \aap, 436, 647

\bibitem[{{Falanga} {et~al.}(2008){Falanga}, {Chenevez}, {Cumming}, {Kuulkers},
  {Trap}, \& {Goldwurm}}]{falanga08}
{Falanga}, M., {Chenevez}, J., {Cumming}, A., {et~al.} 2008, \aap, 484, 43

\bibitem[{{Falanga} {et~al.}(2009){Falanga}, {Cumming}, {Bozzo}, \&
  {Chenevez}}]{falanga09}
{Falanga}, M., {Cumming}, A., {Bozzo}, E., \& {Chenevez}, J. 2009, \aap, 496,
  333

\bibitem[{{Falanga} {et~al.}(2005{\natexlab{b}}){Falanga}, {Kuiper},
  {Poutanen}, {Bonning}, {Hermsen}, {di Salvo}, {Goldoni}, {Goldwurm}, {Shaw},
  \& {Stella}}]{falanga05}
{Falanga}, M., {Kuiper}, L., {Poutanen}, J., {et~al.} 2005{\natexlab{b}}, \aap,
  444, 15

\bibitem[{{Falanga} {et~al.}(2011){Falanga}, {Kuiper}, {Poutanen}, {Galloway},
  {Bonning}, {Bozzo}, {Goldwurm}, {Hermsen}, \& {Stella}}]{falanga11}
{Falanga}, M., {Kuiper}, L., {Poutanen}, J., {et~al.} 2011, \aap, 529, A68

\bibitem[{{Falanga} {et~al.}(2012){Falanga}, {Kuiper}, {Poutanen}, {Galloway},
  {Bozzo}, {Goldwurm}, {Hermsen}, \& {Stella}}]{falanga12}
{Falanga}, M., {Kuiper}, L., {Poutanen}, J., {et~al.} 2012, \aap, 545, A26

\bibitem[{{Falanga} {et~al.}(2007){Falanga}, {Poutanen}, {Bonning}, {Kuiper},
  {Bonnet-Bidaud}, {Goldwurm}, {Hermsen}, \& {Stella}}]{falanga07}
{Falanga}, M., {Poutanen}, J., {Bonning}, E.~W., {et~al.} 2007, \aap, 464, 1069

\bibitem[{{Ferrigno} {et~al.}(2014){Ferrigno}, {Bozzo}, {Papitto}, {Rea},
  {Pavan}, {Campana}, {Wieringa}, {Filipovi{\'c}}, {Falanga}, \&
  {Stella}}]{ferrigno14}
{Ferrigno}, C., {Bozzo}, E., {Papitto}, A., {et~al.} 2014, \aap, 567, A77

\bibitem[{{Fujimoto} {et~al.}(1987){Fujimoto}, {Sztajno}, {Lewin}, \& {van
  Paradijs}}]{fujimoto87}
{Fujimoto}, M.~Y., {Sztajno}, M., {Lewin}, W.~H.~G., \& {van Paradijs}, J.
  1987, \apj, 319, 902

\bibitem[{{Fushiki} \& {Lamb}(1987)}]{fushiki87}
{Fushiki}, I. \& {Lamb}, D.~Q. 1987, \apjl, 323, L55

\bibitem[{{Galloway} \& {Cumming}(2006)}]{galloway06}
{Galloway}, D.~K. \& {Cumming}, A. 2006, \apj, 652, 559

\bibitem[{{Galloway} {et~al.}(2004){Galloway}, {Cumming}, {Kuulkers},
  {Bildsten}, {Chakrabarty}, \& {Rothschild}}]{galloway04}
{Galloway}, D.~K., {Cumming}, A., {Kuulkers}, E., {et~al.} 2004, \apj, 601, 466

\bibitem[{{Gierli{\'n}ski} \& {Zdziarski}(2005)}]{gierlienski05}
{Gierli{\'n}ski}, M. \& {Zdziarski}, A.~A. 2005, \mnras, 363, 1349

\bibitem[{{Gilfanov} {et~al.}(1998){Gilfanov}, {Revnivtsev}, {Sunyaev}, \&
  {Churazov}}]{gilfanov98}
{Gilfanov}, M., {Revnivtsev}, M., {Sunyaev}, R., \& {Churazov}, E. 1998, \aap,
  338, L83

\bibitem[{{Hartman} {et~al.}(2011){Hartman}, {Galloway}, \&
  {Chakrabarty}}]{hartman11}
{Hartman}, J.~M., {Galloway}, D.~K., \& {Chakrabarty}, D. 2011, \apj, 726, 26

\bibitem[{{Heinke} {et~al.}(2015){Heinke}, {Bahramian}, {Degenaar}, \&
  {Wijnands}}]{heinke15}
{Heinke}, C.~O., {Bahramian}, A., {Degenaar}, N., \& {Wijnands}, R. 2015,
  \mnras, 447, 3034

\bibitem[{{Illarionov} \& {Sunyaev}(1975)}]{illarionov75}
{Illarionov}, A.~F. \& {Sunyaev}, R.~A. 1975, \aap, 39, 185

\bibitem[{{in't Zand} {et~al.}(2005){in't Zand}, {Cumming}, {van der Sluys},
  {Verbunt}, \& {Pols}}]{zand05}
{in't Zand}, J.~J.~M., {Cumming}, A., {van der Sluys}, M.~V., {Verbunt}, F., \&
  {Pols}, O.~R. 2005, \aap, 441, 675

\bibitem[{{Jansen} {et~al.}(2001){Jansen}, {Lumb}, {Altieri}, {Clavel}, {Ehle},
  {Erd}, {Gabriel}, {Guainazzi}, {Gondoin}, {Much}, {Munoz}, {Santos},
  {Schartel}, {Texier}, \& {Vacanti}}]{jansen01}
{Jansen}, F., {Lumb}, D., {Altieri}, B., {et~al.} 2001, \aap, 365, L1

\bibitem[{{Kennea} {et~al.}(2013){Kennea}, {Burrows}, {Cummings}, {Markwardt},
  {Gehrels}, {Ukwatta}, {Sbarufatti}, {Malesani}, \& {Mountford}}]{kennea13}
{Kennea}, J.~A., {Burrows}, D.~N., {Cummings}, J.~R., {et~al.} 2013, The
  Astronomer's Telegram, 5354, 1

\bibitem[{{King} \& {Ritter}(1998)}]{king98}
{King}, A.~R. \& {Ritter}, H. 1998, \mnras, 293, L42

\bibitem[{{Krimm} {et~al.}(2013){Krimm}, {Holland}, {Corbet}, {Pearlman},
  {Romano}, {Kennea}, {Bloom}, {Barthelmy}, {Baumgartner}, {Cummings},
  {Gehrels}, {Lien}, {Markwardt}, {Palmer}, {Sakamoto}, {Stamatikos}, \&
  {Ukwatta}}]{krimm13}
{Krimm}, H.~A., {Holland}, S.~T., {Corbet}, R.~H.~D., {et~al.} 2013, \apjs,
  209, 14

\bibitem[{{Kuulkers} {et~al.}(2003){Kuulkers}, {den Hartog}, {in't Zand},
  {Verbunt}, {Harris}, \& {Cocchi}}]{kuulkers03}
{Kuulkers}, E., {den Hartog}, P.~R., {in't Zand}, J.~J.~M., {et~al.} 2003,
  \aap, 399, 663

\bibitem[{{Kuulkers} {et~al.}(2013){Kuulkers}, {Eckert}, {Ferrigno},
  {Chenevez}, {Alfonso-Garzon}, {Beckmann}, {Bird}, {Brandt}, {Del Santo},
  {Domingo}, {Ebisawa}, {Jonker}, {Kretschmar}, {Markwardt}, {Oosterbroek},
  {Paizis}, {Pottschmidt}, {Sanchez-Fernandez}, \& {Wijnands}}]{kuulkers13}
{Kuulkers}, E., {Eckert}, D., {Ferrigno}, C., {et~al.} 2013, The Astronomer's
  Telegram, 5361, 1

\bibitem[{{Kuulkers} {et~al.}(2009){Kuulkers}, {in't Zand}, \&
  {Lasota}}]{kuulkers09}
{Kuulkers}, E., {in't Zand}, J.~J.~M., \& {Lasota}, J.-P. 2009, \aap, 503, 889

\bibitem[{{Kuulkers} {et~al.}(2002){Kuulkers}, {in't Zand}, {van Kerkwijk},
  {Cornelisse}, {Smith}, {Heise}, {Bazzano}, {Cocchi}, {Natalucci}, \&
  {Ubertini}}]{kuulkers02}
{Kuulkers}, E., {in't Zand}, J.~J.~M., {van Kerkwijk}, M.~H., {et~al.} 2002,
  \aap, 382, 503

\bibitem[{{La Parola} {et~al.}(2013){La Parola}, {Segreto}, {Cusumano}, \&
  {Maselli}}]{laparola13}
{La Parola}, V., {Segreto}, A., {Cusumano}, G., \& {Maselli}, A. 2013, The
  Astronomer's Telegram, 5646, 1

\bibitem[{{Lebrun} {et~al.}(2003){Lebrun}, {Leray}, {Lavocat}, {Cr{\'e}tolle},
  {Arqu{\`e}s}, {Blondel}, {Bonnin}, {Bou{\`e}re}, {Cara}, {Chaleil}, {Daly},
  {Desages}, {Dzitko}, {Horeau}, {Laurent}, {Limousin}, {Mathy}, {Mauguen},
  {Meignier}, {Molini{\'e}}, {Poindron}, {Rouger}, {Sauvageon}, \&
  {Tourrette}}]{lebrun03}
{Lebrun}, F., {Leray}, J.~P., {Lavocat}, P., {et~al.} 2003, \aap, 411, L141

\bibitem[{{Lewin} {et~al.}(1993){Lewin}, {van Paradijs}, \& {Taam}}]{lewin93}
{Lewin}, W.~H.~G., {van Paradijs}, J., \& {Taam}, R.~E. 1993, \ssr, 62, 223

\bibitem[{{Linares} {et~al.}(2008){Linares}, {Wijnands}, {van der Klis},
  {Krimm}, {Markwardt}, \& {Chakrabarty}}]{linares08}
{Linares}, M., {Wijnands}, R., {van der Klis}, M., {et~al.} 2008, \apj, 677,
  515

\bibitem[{{Lund} {et~al.}(2003){Lund}, {Budtz-J{\o}rgensen}, {Westergaard},
  {Brandt}, {Rasmussen}, {Hornstrup}, {Oxborrow}, {Chenevez}, {Jensen},
  {Laursen}, {Andersen}, {Mogensen}, {Rasmussen}, {Om{\o}}, {Pedersen},
  {Polny}, {Andersson}, {Andersson}, {K{\"a}m{\"a}r{\"a}inen}, {Vilhu},
  {Huovelin}, {Maisala}, {Morawski}, {Juchnikowski}, {Costa}, {Feroci},
  {Rubini}, {Rapisarda}, {Morelli}, {Carassiti}, {Frontera}, {Pelliciari},
  {Loffredo}, {Mart{\'{\i}}nez N{\'u}{\~n}ez}, {Reglero}, {Velasco}, {Larsson},
  {Svensson}, {Zdziarski}, {Castro-Tirado}, {Attina}, {Goria}, {Giulianelli},
  {Cordero}, {Rezazad}, {Schmidt}, {Carli}, {Gomez}, {Jensen}, {Sarri},
  {Tiemon}, {Orr}, {Much}, {Kretschmar}, \& {Schnopper}}]{lund03}
{Lund}, N., {Budtz-J{\o}rgensen}, C., {Westergaard}, N.~J., {et~al.} 2003,
  \aap, 411, L231

\bibitem[{{Malesani} {et~al.}(2013){Malesani}, {Kennea}, {Burrows}, {Cummings},
  {Markwardt}, {Gehrels}, {Ukwatta}, {Sbarufatti}, {Mountford}, {Krimm},
  {Siegel}, {Chester}, \& {Grupe}}]{malesani13}
{Malesani}, D., {Kennea}, J.~A., {Burrows}, D.~N., {et~al.} 2013, GRB
  Coordinates Network, 15172, 1

\bibitem[{{Molkov} {et~al.}(2005){Molkov}, {Revnivtsev}, {Lutovinov}, \&
  {Sunyaev}}]{molkov05}
{Molkov}, S., {Revnivtsev}, M., {Lutovinov}, A., \& {Sunyaev}, R. 2005, \aap,
  434, 1069

\bibitem[{{Papitto} {et~al.}(2007){Papitto}, {di Salvo}, {Burderi}, {Menna},
  {Lavagetto}, \& {Riggio}}]{papitto07}
{Papitto}, A., {di Salvo}, T., {Burderi}, L., {et~al.} 2007, \mnras, 375, 971

\bibitem[{{Powell} {et~al.}(2007){Powell}, {Haswell}, \& {Falanga}}]{powell07}
{Powell}, C.~R., {Haswell}, C.~A., \& {Falanga}, M. 2007, \mnras, 374, 466

\bibitem[{{Romano} {et~al.}(2006){Romano}, {Campana}, {Chincarini}, {Cummings},
  {Cusumano}, {Holland}, {Mangano}, {Mineo}, {Page}, {Pal'Shin}, {Rol},
  {Sakamoto}, {Zhang}, {Aptekar}, {Barbier}, {Barthelmy}, {Beardmore}, {Boyd},
  {Burrows}, {Capalbi}, {Fenimore}, {Frederiks}, {Gehrels}, {Giommi}, {Goad},
  {Godet}, {Golenetskii}, {Guetta}, {Kennea}, {La Parola}, {Malesani},
  {Marshall}, {Moretti}, {Nousek}, {O'Brien}, {Osborne}, {Perri}, \&
  {Tagliaferri}}]{Romano2006:060124}
{Romano}, P., {Campana}, S., {Chincarini}, G., {et~al.} 2006, \aap, 456, 917

\bibitem[{{Romanova} {et~al.}(2014){Romanova}, {Lovelace}, {Bachetti},
  {Blinova}, {Koldoba}, {Kurosawa}, {Lii}, \& {Ustyugova}}]{romanova14}
{Romanova}, M.~M., {Lovelace}, R.~V.~E., {Bachetti}, M., {et~al.} 2014, in
  European Physical Journal Web of Conferences, Vol.~64, European Physical
  Journal Web of Conferences, 5001

\bibitem[{{Strohmayer} \& {Bildsten}(2006)}]{strohmayer06}
{Strohmayer}, T. \& {Bildsten}, L. 2006, {New views of thermonuclear bursts},
  ed. W.~H.~G. {Lewin} \& M.~{van der Klis}, 113--156

\bibitem[{{Ubertini} {et~al.}(2003){Ubertini}, {Lebrun}, {Di Cocco}, {Bazzano},
  {Bird}, {Broenstad}, {Goldwurm}, {La Rosa}, {Labanti}, {Laurent}, {Mirabel},
  {Quadrini}, {Ramsey}, {Reglero}, {Sabau}, {Sacco}, {Staubert}, {Vigroux},
  {Weisskopf}, \& {Zdziarski}}]{ubertini03}
{Ubertini}, P., {Lebrun}, F., {Di Cocco}, G., {et~al.} 2003, \aap, 411, L131

\bibitem[{{Vaughan} {et~al.}(1994){Vaughan}, {van der Klis}, {Wood}, {Norris},
  {Hertz}, {Michelson}, {van Paradijs}, {Lewin}, {Mitsuda}, \&
  {Penninx}}]{vaughan94}
{Vaughan}, B.~A., {van der Klis}, M., {Wood}, K.~S., {et~al.} 1994, \apj, 435,
  362

\bibitem[{{Vaughan} {et~al.}(2006){Vaughan}, {Goad}, {Beardmore}, {O'Brien},
  {Osborne}, {Page}, {Barthelmy}, {Burrows}, {Campana}, {Cannizzo}, {Capalbi},
  {Chincarini}, {Cummings}, {Cusumano}, {Giommi}, {Godet}, {Hill}, {Kobayashi},
  {Kumar}, {La Parola}, {Levan}, {Mangano}, {M{\'e}sz{\'a}ros}, {Moretti},
  {Morris}, {Nousek}, {Pagani}, {Palmer}, {Racusin}, {Romano}, {Tagliaferri},
  {Zhang}, \& {Gehrels}}]{vaughan2006:050315}
{Vaughan}, S., {Goad}, M.~R., {Beardmore}, A.~P., {et~al.} 2006, \apj, 638, 920

\end{thebibliography}

\end{document}